# Decision Tree for Protein Biomarker Selection for Clinical Applications

Katharina Waury[1]

[1]Department of Computer Science, Vrije Universiteit Amsterdam, Amsterdam, The Netherlands, k.waury@vu.nl

**Abstract**

Discovery of novel protein biomarkers for clinical applications is an active research field across a manifold of diseases. Despite some successes and progress, the biomarker development pipeline still frequently ends in failure as biomarker candidates cannot be validated or translated to immunoassays. Selection of strong disease biomarker candidates that further constitute suitable targets for antibody binding in immunoassays is thus important. This essential selection step can be supported and rationalized using bioinformatics tools such as protein databases. Here, I present a workflow in the form of decision trees to computationally investigate biomarker candidates and their available affinity reagents in depth. This analysis can identify the most promising biomarker candidates for assay development while minimal time and effort is required.

## 1. Introduction

Biomarkers are an important cornerstone of clinical research and practice as they support patient diagnosis, disease monitoring and the implementation of clinical drug trials (1). Especially protein biomarkers have received increased interest and the use of fluid protein markers in clinical practice is now strongly established for multiple diseases (2,3). Nevertheless, many needs in the medical field are still unmet and the discovery and development of protein biomarkers is an ongoing and active research field (4,5). Novel candidates are constantly reported; often based on the results of high-throughput proteomics studies of body fluids that compare the protein concentrations between diseased and healthy groups to identify proteins with meaningful differences in concentration (5). Despite the plethora of candidate proteins, rarely any biomarkers reach the clinical practice, as the biomarkers pipeline is challenging and failure prone after the discovery phase (6). One especially prevalent bottleneck is the technology-translation gap between the technologies used for discovery and the ones used for validation and clinical implementation (4,7). While not suitable for large-scale biomarker discovery, immunoassays are essential at a later stage to use biomarkers in routine clinical practice (8). Thus, to facilitate clinical biomarkers, it is important to establish that a protein is: 1) a reliable marker for the disease or biological process of interest, and 2) a feasible target molecule in an immunoassay. Examination and confirmation of these two properties using laboratory experiments is time-consuming and costly. Considering that biomarker discovery proteomics studies often lead to a long list of candidates, it is not feasible to test all proteins thoroughly and a selection of the most promising candidates is required (7).

Bioinformatics resources have been suggested to be a suitable approach to support this selection process (7,9). Importantly, this step requires minimal effort while potentially saving a vast amount of energy and costs otherwise spent on unsuccessful assay development. The inclusion of information provided in online databases and other resources would allow to corroborate two main attributes for an investigated protein:

1. The protein is a strong biomarker candidate for a disease of interest based on available information including its known disease associations, function, and interaction.
2. The protein is a suitable target for antibody binding to facilitate its use in clinical immunoassays considering the available antibody reagents and the antibody-binding region within the protein.

This chapter introduces biomarker researchers to the available resources and illustrates how to utilize the provided information for biomarker candidate selection. First, all proposed online resources for biomarker candidate screening are summarized in the following *Resources* section. Next, in the *Methods* section a workflow in the form of two decision trees is presented. Each decision tree is suited to confirm one of the two aforementioned desired attributes of clinical biomarkers using computational resources.

## 2. Resources

The suggested resources for a biomarker suitability survey are described shortly below. All included tools are free to use, are thoroughly maintained and updated, and are meant to be easily usable by non-bioinformatics researchers. All tools are summarized in **Table 1** which also provides the current web address and the year of their last major update (see **Note 1**). The suggested use of these tools in the context of biomarker candidate discovery and selection is explained in the following section.

### a. UniProt

The UniProt Knowledge Database (10) is the most comprehensive data resource for proteins currently available. Especially for human proteins a wide range of high-quality, manually curated annotations is collected and presented in a consistent manner. The database is constantly updated and new cross-references to other major bioinformatics tools are included. Important annotations for each protein include gene ontology terms, subcellular location, post-translational modifications (PTMs) and the sequences of the canonical protein and potentially other isoforms.

### b. MarkerDB

The MarkerDB database (11) collects all known clinical and pre-clinical biomarkers in one resource. It currently contains 142 protein biomarkers for which information is provided on the associated condition and the biomarker type. It also provides the currently established biomarker concentration cutoffs in biological fluids used in clinical practice.

### c. DisGeNet

Gene-disease associations are compiled in DisGeNet (12) based on scientific literature, curation by experts and high-throughput studies, e.g., GWAS. Disease associations at the gene level can indicate a role of the linked protein and thus this information can be suitable to identify protein biomarkers. The collection of relevant publications provides an estimate on how long and how well a specific disease association has already been studied by the scientific community.

Table 1: Overview of bioinformatics databases for biomarker candidate selection. Cont. – Continuously updated, HBFP – Human Body Fluid Proteome, HPA – Human Protein Atlas, PPD – Plasma Protein Database, PTM – Post-translational modification

| Database | Information | URL | Ref. | Last update |
|---|---|---|---|---|
| UniProt | Various | https://www.uniprot.org/ | (10) | Cont. |
| MarkerDB | Biomarkers-disease-associations | https://markerdb.ca/ | | 2020 |
| DisGeNet | Gene-disease-associations, biomarker-disease-associations | https://www.disgenet.org/; https://www.disgenet.org/biomarkers/ | (12), (13) | 2020, 2022 |
| STRING | Protein-protein-interactions | https://string-db.org/ | (14) | 2023 |
| HBFP | Body fluid proteins | https://bmbl.bmi.osumc.edu/HBFP/ | (15) | 2021 |
| PPD | Plasma proteins | http://www.plasmaproteomedatabase.org/ | (16) | Unknown |
| HPA | Various | https://www.proteinatlas.org/ | (17) | Cont. |
| ProteomicsDB | Various | https://www.proteomicsdb.org/ | (18) | Cont. |
| PaxDb | Protein abundance | https://www.pax-db.org/ | (19) | 2023 |
| SheddomeDB | Ectodomain shedding proteins | https://bal.lab.nycu.edu.tw/sheddomedb/ | (20) | 2023 |
| Vesiclepedia | Extracellular vesicle cargo proteins | http://microvesicles.org/ | (21) | 2018 |
| Antibodypedia | Antibody validation | https://www.antibodypedia.com/ | (22) | Cont. |
| CiteAb | Antibody validation | https://www.citeab.com/ | (23) | Cont. |
| iPTMnet | PTMs | https://research.bioinformatics.udel.edu/iptmnet/ | (24) | 2022 |
| InterPro | Domains | https://www.ebi.ac.uk/interpro/ | (25) | Cont. |
| AlphaFold Database | Predicted protein structures | https://alphafold.ebi.ac.uk/ | (26) | 2023 |
| MobiDB | Disorder, interacting residues | https://mobidb.bio.unipd.it/ | (27) | 2022 |

Recently, the DisGeNet database was expanded to include the Clinical Biomarker App (13). Here, currently over 3000 proteins are presented regarding the diseases and clinical trials they are utilized for as a biomarker.

d. STRING

STRING (14) is a protein-protein-interaction database that derives its information from curations, scientific literature by text mining as well as computational predictions (see **Note 2**). As a large number of organisms are included, it is important to select the human proteome when searching for interactions of a protein of interest. The interactions are visualized as a graph network which also includes the interplay between a protein's interaction partners. Every interaction is annotated with a confidence score based on the available evidence.

e. HBFP

The Human Body Fluid Proteome (HBFP) database (15) currently contains over 11,000 proteins detected in 17 body fluids. For many less well researched fluids, this might be the single resource available collecting the results of multiple proteomics studies. Every protein entry provides a confidence score per body fluid and the references of the studies are included in the database.

f. Plasma Proteome Database

The Plasma Proteome Database (16) is a project initiated by the Human Proteome Project. It is a collection of over 10,000 proteins that have been reported in proteomics studies of human plasma. Entries report on the available experimental evidence for each protein and if provided, the measured protein concentrations.

g. Human Protein Atlas

The Human Protein Atlas (17) is a constantly expanding project that aims to map the entire human proteome regarding various aspects. Presently, it contains sections on tissue and cell type expression, subcellular location and several specific atlases, e.g., of the brain proteome and the blood proteome. Different technologies were used to collect the information, such as RNA sequencing, mass spectrometry and immunoassays.

h. ProteomicsDB

The ProteomicsDB (18) is a mainly mass spectrometry-focused resource which is continuously expanded and updated. It contains a wide range of information for the human proteome in a standardized manner. Importantly, the database visualizes tissue-specific protein expression for every protein and provides a feature viewer which includes PTM, domain and structure annotations.

i. PaxDb

The protein abundance database PaxDb (19) collects protein quantification datasets to provide a unified resource of protein abundances across different tissues. Integrated datasets in the database provide very high coverage of the human proteome and tissue-specific datasets are also available.

j. SheddomeDB

SheddomeDB (20) is a recently updated database of known ectodomain shedding proteins based on the curation of scientific literature. Albeit containing less than 500 entries across multiple organisms, SheddomeDB is a valuable resource as ectodomain shedding is an underrepresented annotation despite its importance for biomarker research.

k. Vesiclepedia

Vesiclepedia (21) is a resource that collects studies of extracellular vesicles and the identified cargo, including proteins. Each protein entry contains information about the studies which confirmed its association with extracellular vesicles. Details on study design, sample material and vesicle type are also provided. Note that the database was last updated in 2018 and hence does not contain study results of recent years.

l. Antibodypedia

Antibodypedia (22) is a resource to search for antibodies and the results of their validation experiments. Data can be provided by antibody suppliers as well as researchers but is reviewed before inclusion. Importantly, validation data can be searched for a specific application which allows researchers to find antibodies most suitable for their intended use.

m. CiteAb

The CiteAb database (23) ranks included antibodies based on their references in scientific publications. This approach lets researchers discover the reagents that are most trusted by the research community and provides a simple strategy to find all the relevant literature which can also be filtered regarding the desired application type.

n. iPTMnet

The database iPTMnet (24) contains detailed information on known protein modification sites. These PTM sites are visualized across variants and isoforms and are annotated with confidence score based on the available evidence. As iPTMnet integrates knowledge from several other databases it contains a comprehensive overview of current information on PTMs.

o. InterPro

InterPro (25) integrates data from several other resources. The database classifies proteins into families based on annotations regarding domains and motifs and visualizes those along the protein sequence. This information is provided across all isoforms of a proteins. InterPro can also be used to discover proteins similar to the protein of interest.

p. BLAST

The Basic Local Alignment Search Tool (BLAST) (28) can identify regions of high sequence similarity both at a nucleotide and protein level. The tool can be used to find similar proteins within a proteome set, such as the Swiss-Prot human proteome, for a protein sequence or sequence fragment of interest.

q. AlphaFold Database

The AlphaFold database (26) is the largest collection of high-quality protein structures that were predicted by a state-of-the-art machine learning model (29) (see **Note 2**). Almost the entire human proteome is covered. For each protein entry the predicted structure is displayed which includes information regarding the confidence of the predictions. The tool menu allows to easily highlight protein regions of interest and both the structure models and screenshots thereof can be easily downloaded.

r. MobiDB

MobiDB (27) is a database focused on protein structure and binding annotations. It provides information on disordered and structured regions as well as linear interacting peptides within a protein. These properties are visualized along the sequence which facilitates identification of longer stretches of disordered coil regions within a protein. The annotations are based both on experimental observations and machine learning predictions (see **Note 2**).

## 3. Methods

Here, a protocol is presented to facilitate the rationalized selection of biomarker candidates using available online bioinformatics tools, e.g., curated protein databases. Decision trees illustrate and summarize in which explicit way these tools can be utilized to select proteins that are a strong disease biomarker candidate (**Figure 1**), and to select proteins that are a suitable target in antibody-based immunoassays (**Figure 2**). The decision trees show for each protocol step which supportive or discouraging conclusions regarding a protein's suitability as a biomarker can be made. Supportive results suggest to favor a protein, while discouraging results might indicate to exclude a protein from further biomarker development or suggest that special care needs to be applied to deal with a potential issue. Protocol steps 1-6 are part of the first decision tree (**Figure 1**); protocol steps 7-12 are included in the second decision tree (**Figure 2**).

### 3.1. Define the Protein Identifier

A first important step to examine a protein using online bioinformatics resources is to identify its associated identifier in UniProt (10) (see **Note 3**). The UniProt ID (also known as accession number) is highly preserved and is often required to search other protein database entries. Additionally, because of the many cross-references to other databases, UniProt provides a suitable starting point to examine online information for a protein.

### 3.2. Search for Biomarker-Disease Associations

Any existing knowledge of a biomarker candidate's association on the protein or gene level with pathologies should be investigated. Next to a conventional search in databases of scientific literature, e.g., PubMed or Google Scholar, specific databases for biomarkers and disease associations, i.e., MarkerDB (11) and DisGeNet (12,13), can be explored for a more concise overview. A protein's association with a disease of similar pathology or the same involved pathways as the disease of interest might constitute a strong biomarker candidate. On the other hand, associations with other diseases than the one of interest might limit the disease specificity of a biomarker, especially for diseases which are known to co-occur substantially.

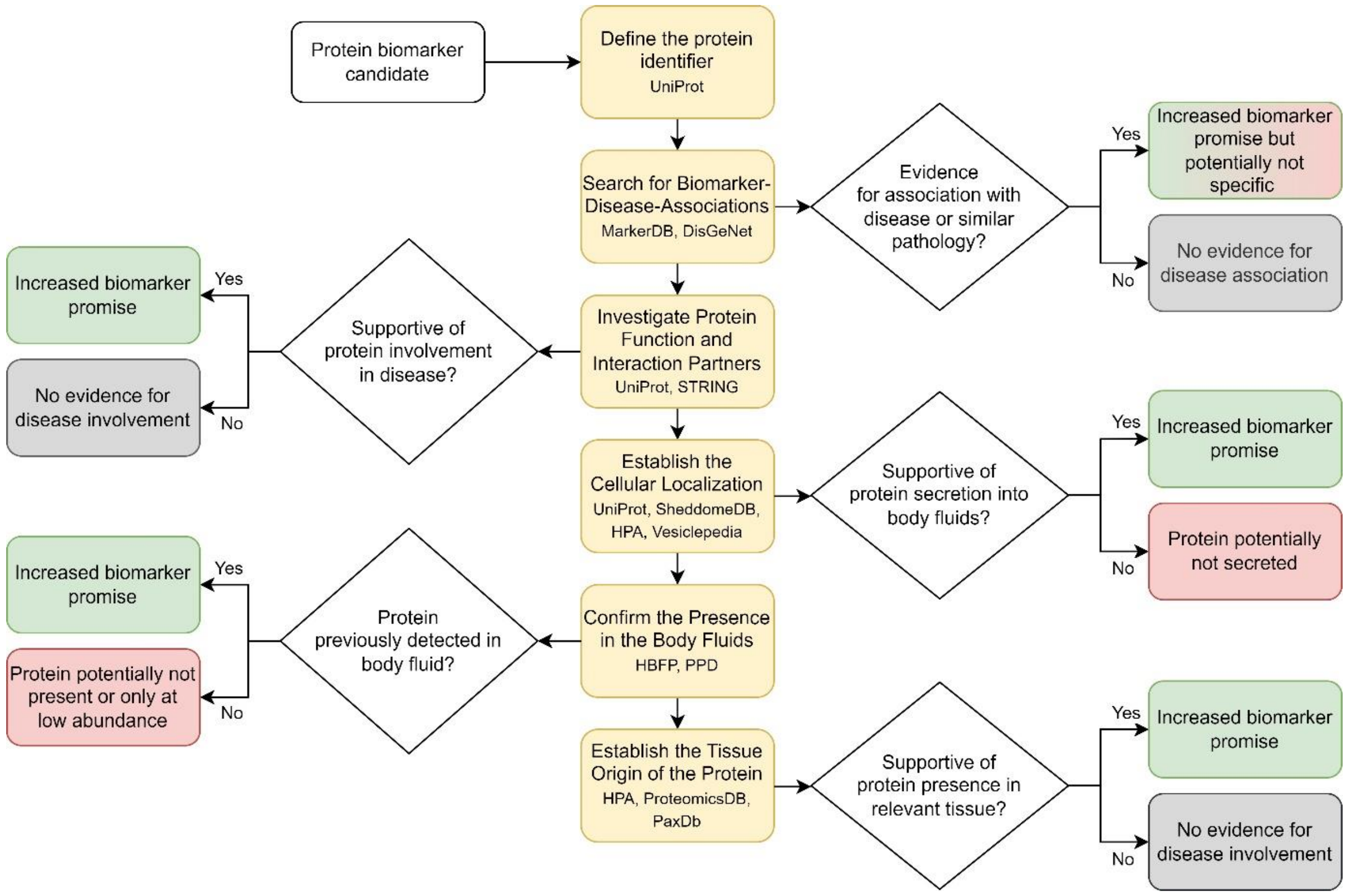

Figure 1: Decision tree to establish a protein candidate as a marker of disease. Steps to evaluate the suitability of a protein as a specific disease marker are shown as a decision tree flow chart. Yellow boxes indicate the steps to follow as well as the advised tools to use. Each step attempts to answer the indicated question. The outcome boxes are colored according to the implied assessment: supportive (green), opposing (red), or neutral (grey). HBFP – Human Body Fluid Proteome, HPA – Human Protein Atlas, PPD – Plasma Protein Database

### 3.3. Investigate Protein Function and Interaction Partners

A protein's functions and interactions increase the understanding of the mechanisms through which a protein is potentially involved in disease pathology. Protein function can be examined on UniProt (10) in the form of gene ontology terms. Involvement of a protein in a pathway that is affected by the investigated pathology bolsters its role as a biomarker. Protein-protein-interaction partners listed in the STRING database (14) that are known to be involved in the disease of interest also strengthen a protein's biomarker capacity.

### 3.4. Establish the Cellular Localization of the Protein

The subcellular localization of a protein might also be of interest as it informs on the protein's role within the cell. Further, it has been shown that the location within the cell influences a protein's probability to be secreted to the body fluids (30); a protein found at the membrane or extracellular space might thus be more suitable as a fluid biomarker. UniProt (10) and the Human Protein Atlas (17) can be explored for this information. Ectodomains which can be shed into the body fluids as well as association with extracellular vesicles are relevant properties that might affect a protein's biomarker suitability (30). Both circumstances can lead to a protein being present in a body fluid. Specialized databases, SheddomeDB (20) and Vesiclepedia (21), include this type of information.

### 3.5. Confirm the Presence of the Protein in the Body Fluids

Protein biomarkers are preferably measured in body fluids because of the vast amount of secreted proteins present and the non-invasive sample extraction (5). Important body fluids for biomarker discovery include plasma, urine, and cerebrospinal fluid (15). The detection of a biomarker candidate in the body fluid of interest can be confirmed in the Human Body Fluid Proteome database (15). Proteins specifically detected in plasma can further be confirmed in the Plasma Proteome Database (16).

### 3.6. Establish the Tissue Origin of the Protein

It is beneficial to establish the protein's likely tissue origin if it is measured in a body fluid. A protein that is highly and specifically expressed in the tissue affected by the pathology of interest has strong biomarker potential. For example, the brain is of the highest interest regarding the search for novel biomarker for dementia such as Alzheimer's Disease. Ubiquitous expression convolutes the determination of a biomarker's origin. If the biomarker candidate is very lowly expressed in the tissue of interest, it might only be detectable in body fluids with ultrasensitive methods (31). Information regarding protein expression and abundance in the different human tissues can be checked in the databases of the Human Protein Atlas (17), ProteomicsDB (18) and PaxDb (19).

### 3.7. Check the Availability of Suitable Affinity Reagents

After strong candidates for disease protein biomarkers have been identified following steps 1 to 6, their suitability as a target molecule for detection by antibodies in an immunoassay set up, e.g., an enzyme-linked immunosorbent assay (ELISA), is evaluated in the last six protocol

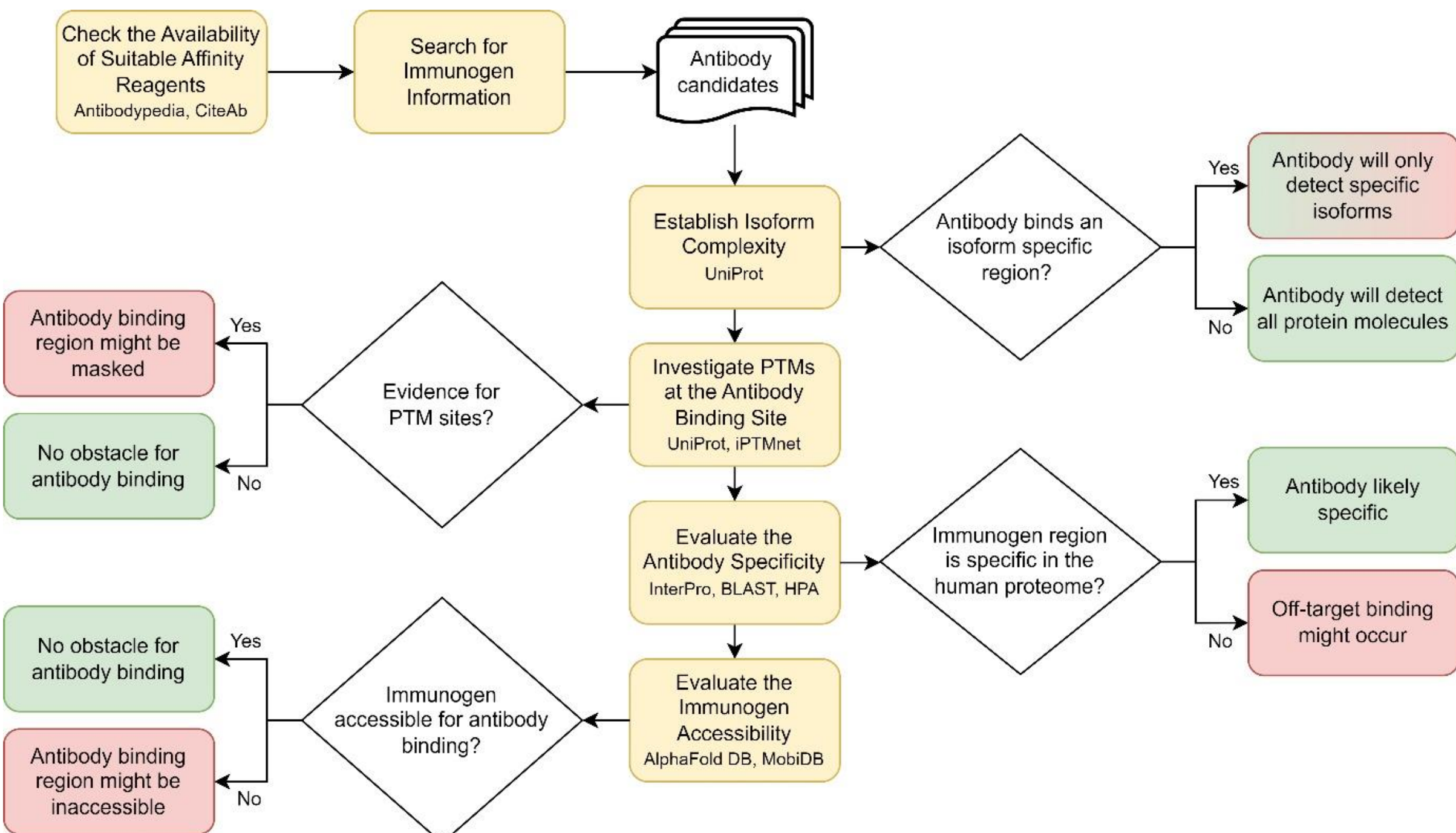

Figure 2: Decision tree to establish a protein as a suitable assay target. Steps to evaluate the suitability of a protein as an immunoassay target for antibody binding are shown as a decision tree flow chart. Yellow boxes indicate the steps to follow as well as the advised tools to use. Each step attempts to answer the indicated question. The outcome boxes are colored according to the implied assessment: supportive (green) or opposing (red). BLAST – Basic Local Alignment Search Tool, HPA – Human Protein Atlas, PTM – Post-translational modification

steps. Note that these steps should be done individually for each protein that constitutes a strong biomarker candidate according to the previous analysis.

Firstly, the availability of affinity reagents, i.e., commercial antibodies or immunoassay kits, should be reviewed for each protein. The antibody databases Antibodypedia (22) and CiteAb (23) as well as other online resources, e.g., Antibodies Online (https://www.antibodies-online.com), Linscott's Directory (https://www.linscottsdirectory.com/) and Biocompare (https://www.biocompare.com/), can be searched. Attention should be paid to information regarding each antibody's clonality, the host organism, the immunogen used for animal immunization, and for which application the antibody was already validated (32). This step leads to a list of potential antibodies for immunoassay development.

### 3.8. Search for Immunogen Information

It should be checked for which antibodies information on the utilized immunogen is available as this allows further assessment of the antibody's suitability for the desired immunoassay application. If the immunogen is not disclosed on the supplier's webpage, requesting the information directly from the supplier is possible but might not be successful (32).

### 3.9. Establish Isoform Complexity

It is important to consider if an antibody will only detect a subset of isoforms of a biomarker candidate. If the immunogen of a considered antibody lies in a region affected by alternative splicing, the isoforms bound and detected by this antibody are limited. This is an advantage if only a specific isoform of a biomarker is desired to be detected but is detrimental if all protein molecules should be measured in the assay. UniProt (10) provides all known isoforms for a protein.

### 3.10. Investigate Post-Translational Modifications at the Antibody Binding Site

PTMs within the binding region of an antibody could mask the recognition site and thus hinder the biomarker-antibody interaction required for a successful assay. UniProt (10) provides location-specific information on known PTMs. Specified PTM databases, e.g., iPTMnet (24), often contain additional knowledge and their combined annotations will thus provide the most complete picture. PTM-specific antibodies exist and are in certain cases specifically required.

### 3.11. Evaluate the Antibody Specificity

Off-target binding of commercial antibodies is a consistently reported issue (33). As validation of antibody specificity in complex fluids is a strenuous process, it is helpful to exclude likely unspecific antibody recognition sites beforehand using bioinformatics tools. If the immunogen is an entire domain of the protein of interest, InterPro (25) annotations can be checked to identify how common this domain is across the human proteome. If a shorter immunogenic peptide was utilized as the immunogen, the specificity of this sequence can be evaluated using BLAST (28). The Human Protein Atlas (17) also provides information on the maximum sequence identify between human proteins on the antibody page of each protein.

### 3.12. Evaluate the Immunogen Accessibility

If an immunogenic peptide was used for immunization, there is no guarantee that it will be accessible to the antibody within the full-length, folded protein structure. While this is less of an issue in applications that contain a denaturation step, e.g., western blots, this can hinder antibody binding in applications detecting the native protein, e.g., ELISAs (34). If the immunogenic peptide is disclosed, its location within the folded protein structure can be inspected to evaluate its accessibility for antibody binding. The AlphaFold Database (26) can be accessed to visualize the immunogen region within the predicted structure. An immunogen located at the surface and preferably in a disordered coil region might be preferred over a peptide positioned within the core of the protein and thus not accessible for antibody binding in an immunoassay application. MobiDB (27) can be accessed to identify if an immunogen is located within a region annotated as disordered. Higher disorder implies the protein is in an unfolded coil structure which is preferred for antibodies that were raised against immunogenic peptides.

## 4. Notes

4.1. Public bioinformatics databases differ in the extent to which they are updated and curated. The most trustworthy resources are maintained or supported by major research institutes and organizations, including ELIXIR, the European Bioinformatics Institute (EMBL-EBI) or the National Institute of Health (NIH). Involved organizations are often displayed by banners on the homepage which are a strong indicator that a resource contains accurate and up-to-date information. **Table 1** informs on the last known update for all included tools.

4.2. When using the results of prediction tools instead of experimentally validated annotations, caution regarding the reliability of this type of information is important. While extremely helpful, machine learning-based predictors are never fully accurate, and the performance of an included tool should always be contemplated. It is advised to research the general performance of prediction tools within a specific field, e.g., protein structure prediction, to assess how trustworthy these are.

4.3. If UniProt is searched by protein or gene name, frequently several protein entries will be returned. To identify the correct entry for the protein of interest, it is important to filter for human proteins and preferably for Swiss-Prot entries. These entries are manually and extensively curated in comparison to the TrEMBL section of UniProt. For the human proteome it should be expected to always find a Swiss-Prot entry as these proteins are well researched and of high importance to the research community.

## 5. Acknowledgements

This work is supported by the European Union's Horizon 2020 research and innovation programme under the Marie Skłodowska-Curie grant agreement No 860197, the MIRIADE project. The author would like to acknowledge the fellow researchers from the MIRIADE consortium who provided valuable insights and interesting discussions regarding fluid biomarker development over the years.